\documentstyle[12pt,aaspp4]{article}
\begin{document}

\title{Near-Infrared Spectroscopy of Photodissociation Regions: The Orion Bar
and Orion~S}
\author{K. L. Luhman\altaffilmark{1}, C. W. Engelbracht\altaffilmark{1},
and M. L. Luhman\altaffilmark{2,3}}

\affil{kluhman@as.arizona.edu, chad@as.arizona.edu, luhman@irfp1.nrl.navy.mil}

\altaffiltext{1}{Steward Observatory, University of Arizona, Tucson, AZ 85721.}

\altaffiltext{2}{Naval Research Laboratory, Remote Sensing Division, Code 7217,
Washington, DC 20375}

\altaffiltext{3}{NRC-NRL Research Associate}

\begin{abstract}

We have obtained moderate-resolution ($\rm R\sim3000$) spectra of the Orion bar 
and Orion~S regions at J (1.25~\micron), H (1.64~\micron), and K (2.2~\micron).
Towards the bar, the observations reveal a large
number of H$_{2}$ emission lines that, when compared to model predictions 
of Draine \& Bertoldi (1996), are indicative of a high-density
photodissociation region (PDR) (n$_{\rm H}=10^6$~cm$^{-3}$, $\chi=10^5$,
T$_{0}=1000$~K) rather than shocked material. Behind the bar into the 
molecular cloud, the H$_{2}$ spectrum again matches well with that predicted
for a dense PDR (n$_{\rm H}=10^6$~cm$^{-3}$) but with a lower temperature 
(T$_{0}=500$~K) and UV field strength ($\chi=10^4$).  The H$_{2}$ spectrum 
and stratification of near-IR emission lines (O~I, H~I, [Fe~II], [Fe~III], 
H$_{2}$) near Orion~S imply the presence of a dense PDR with an inclined 
geometry in this region as well (n$_{\rm H}=10^6$~cm$^{-3}$, $\chi=10^5$, 
T$_{0}=1500$~K).  The extinction measurements towards the bar 
(A$_{\rm K}\sim2.6$) and Orion~S (A$_{\rm K}\sim2.1$) H$_{2}$ emission regions
are much larger than expected from either face-on
(A$_{\rm K}\sim0.1$) or edge-on (A$_{\rm K}\sim1$) homogeneous PDRs,
indicating that clumps may significantly affect the structure of the PDRs. 

In addition, we have observed the strongest $\sim30$ near-IR He~I emission 
lines, many of which have not been detected previously.  There is good 
agreement between most observed and theoretical He~I line ratios, while a few 
transitions with upper levels of $\rm n^3P$ (particularly $\rm 4^3P-3^3S$
1.2531~\micron) are enhanced over strengths expected from collisional 
excitation. This effect is possibly due to opacity in the UV series
$\rm n^{3}P-2^{3}S$.  
We also detect several near-IR [Fe~II] and [Fe~III] transitions with line 
ratios indicative of low densities (n$_{\rm e}\sim10^3$-10$^4$~cm$^{-3}$),
whereas recent observations of optical [Fe~II] emission imply
the presence of high-density gas (n$_{\rm e}\sim10^6$~cm$^{-3}$).
These results are consistent with a model in which high-density, 
partially-ionized gas is the source of the iron transitions
observed in the optical while low-density, fully-ionized material is
responsible for the near-IR emission lines.

\end{abstract}

\keywords{infrared: ISM: lines and bands --- ISM: individual (Orion Nebula) 
--- ISM: molecules --- ISM: structure}

\section{Introduction}

The Orion Nebula is an excellent laboratory for the study of various phenomena
occurring in the interstellar medium. Within this region, the Orion bright bar
has proven especially useful in understanding the detailed structure of
photodissociation regions (PDRs), which have become a subject of intensive
theoretical modeling in recent years (Tielens \& Hollenbach 1985; Burton,
Hollenbach, \& Tielens 1990; Sternberg \& Dalgarno 1995; Draine \& Bertoldi
1996, hereafter DB96) with particular emphasis on the ultraviolet (UV) 
excitation of H$_{2}$ and the subsequent fluorescent infrared (IR) 
emission-line spectra (Black \& van Dishoeck 1987; Sternberg \& Dalgarno 1989).
As a relatively nearby ($\sim450$~pc) and apparently edge-on PDR, the
chemical and thermal stratification within the Orion bar has been well-resolved
and compared extensively to model predictions (Tielens et 
al.\ 1993; Omodaka et al.\ 1994; Tauber et al.\ 1994; Hogerheijde, Jansen, 
\& van Dishoeck 1995; Jansen et al.\ 1995; Usuda et al.\ 1996; van der Werf et
al.\ 1996).

Near-IR spectroscopy of PDR emission features, particularly H$_2$
1-0~S(1) and 2-1~S(1), has been performed towards the Orion bar (Hayashi
et al.\ 1985; Burton et al.\ 1990; Sellgren, Tokunaga, \& Nakada 1990;
Marconi et al.\ 1997), but
with recent advances in near-IR spectrometers we can now reveal the entire IR
spectrum of the Orion PDRs in much greater detail than previously possible. 
For instance, in H- and K-band spectra of the planetary nebula Hubble~12,
which is a source of pure H$_{2}$ fluorescence arising in low-density gas
(n$_{\rm H}\leq10^4$-10$^5$~cm$^{-3}$), Luhman \& Rieke (1996) found
excellent agreement between the observed line ratios of more than 30 H$_2$ 
transitions and the values predicted by theoretical models (Black \& van
Dishoeck 1987). In addition to H$_2$, these spectra of Hubble~12 included many
transitions in He~I, [Fe~II], and [Fe~III], providing tests of the model near-IR
spectra of these species. To test the PDR model spectra
under the physical conditions (temperature, UV field strength, 
density) present in Orion, we have obtained moderate-resolution
($\rm R\sim3000$) J-, H-, and K-band long-slit spectra of the Orion bar and
a bright emission region $\sim40\arcsec$ southwest of the Trapezium
stars, commonly known as Orion~S (Ziurys et al.\ 1981).
These observations provide conclusive evidence for the PDR origin (rather
than shocks) of the H$_{2}$ emission in each of these regions and demonstrate
significant departures in the observed chemical structure from that predicted
by models of homogeneous PDRs. 

\section{Observations}
\label{sec:obs}

On 1995 December 8-10, we performed spectroscopy at J (1.25~\micron),
H (1.64~\micron), and K (2.2~\micron) towards the Orion bright bar and Orion~S
region using the near-IR long-slit spectrometer FSpec (Williams et al.\ 1993) 
at the Steward 1.55~m Bigelow Reflector.  Using a $128\times128$ IR camera as
an off-the-slit guider, we obtained H-band images at each slit position as
a check of the exact coordinates of our apertures. For Orion~S, the 
$3\farcs5\times130\arcsec$ slit ($=2\times75$ pixels) was aligned east-west
and centered at $\alpha=5^{\rm h}35^{\rm m}16\fs8$,
$\delta=-5\arcdeg23\arcmin59\arcsec$ (2000).  In the second set of 
observations, the slit was perpendicular to the bar ($\rm PA=135\arcdeg$) and
centered at 
$\alpha=5^{\rm h}35^{\rm m}20\fs3$, $\delta=-5\arcdeg25\arcmin06\arcsec$ (2000).
One grating setting in J, H, and K covered 0.03, 0.1, and 0.1~\micron\ with
a two-pixel resolution of $\lambda/\Delta\lambda=3200$, 2400, and 3200,
respectively, providing full coverage of these bands with a total of 16 grating
settings.  For each grating setting, we obtained four exposures of 60~s on 
the source in addition to four exposures 5$\arcmin$ to the west to provide a
measurement of the sky background.  We then observed a telluric standard
star by stepping it through six positions along the slit.  Due to poor
weather conditions while obtaining data longward of 2.3~\micron, we
repeated the K-band observations at the 2.3~m Bok Reflector on Kitt Peak
during 1996 October 28 and 29.  The slit was $2\farcs4\times90\arcsec$
($=2\times75$ pixels) at this telescope, but otherwise these observations
were identical to the previous ones.

IRAF routines and customized scripts within this environment were used to
reduce the data. After dark subtraction and flat-fielding, off-source images
were subtracted from on-source data to remove sky emission.  For a given
grating setting, four sky-subtracted images were combined, during which most 
transient bad pixels (e.g., cosmic rays) were rejected. Each combined image 
was then divided by the spectrum of the standard star to correct for telluric 
absorption (see \S~\ref{sec:telluric}).  The resulting image at each
grating setting, consisting of 256 pixels along the dispersion axis and
75 pixels in the spatial direction, was wavelength calibrated using OH 
airglow lines and Ne-Kr lamp spectra.  To correct for slight guiding errors,
we used the stars falling
in the slit and the spatial variation of the hydrogen lines to align
images from consecutive grating settings, which were then combined into
one image spanning J (1.25-1.34~\micron), H (1.45-1.75~\micron), and 
K (2.0-2.45~\micron). The spatial offsets were always less than 2 pixels. 
Since each grating setting overlapped by 50\% with its neighboring spectrum
(e.g., 2.04-2.14~\micron, 2.09-2.19~\micron, etc.), virtually every
emission line appeared in two grating settings. Consequently, we were able
to achieve accurate relative calibration ($<5$\%) across each band.
The telluric standard at H was a G star (HR~2007), the stellar spectral
slope and absorption features of which were removed from the Orion data
by applying the solar spectrum, as discussed by Maiolino, Rieke, \& Rieke
(1996). The Orion data at J and K were multiplied by a blackbody of 
$\rm T=9900$~K to correct for the slope introduced by the A-star telluric 
standard used in these bands (HR~1489 with stellar Pa$\beta$ and Br$\gamma$
absorption removed artificially).

\section{Results}

\subsection{Extraction of Spectra and Flux Calibration}
\label{sec:extract}

\begin{figure}
\plotfiddle{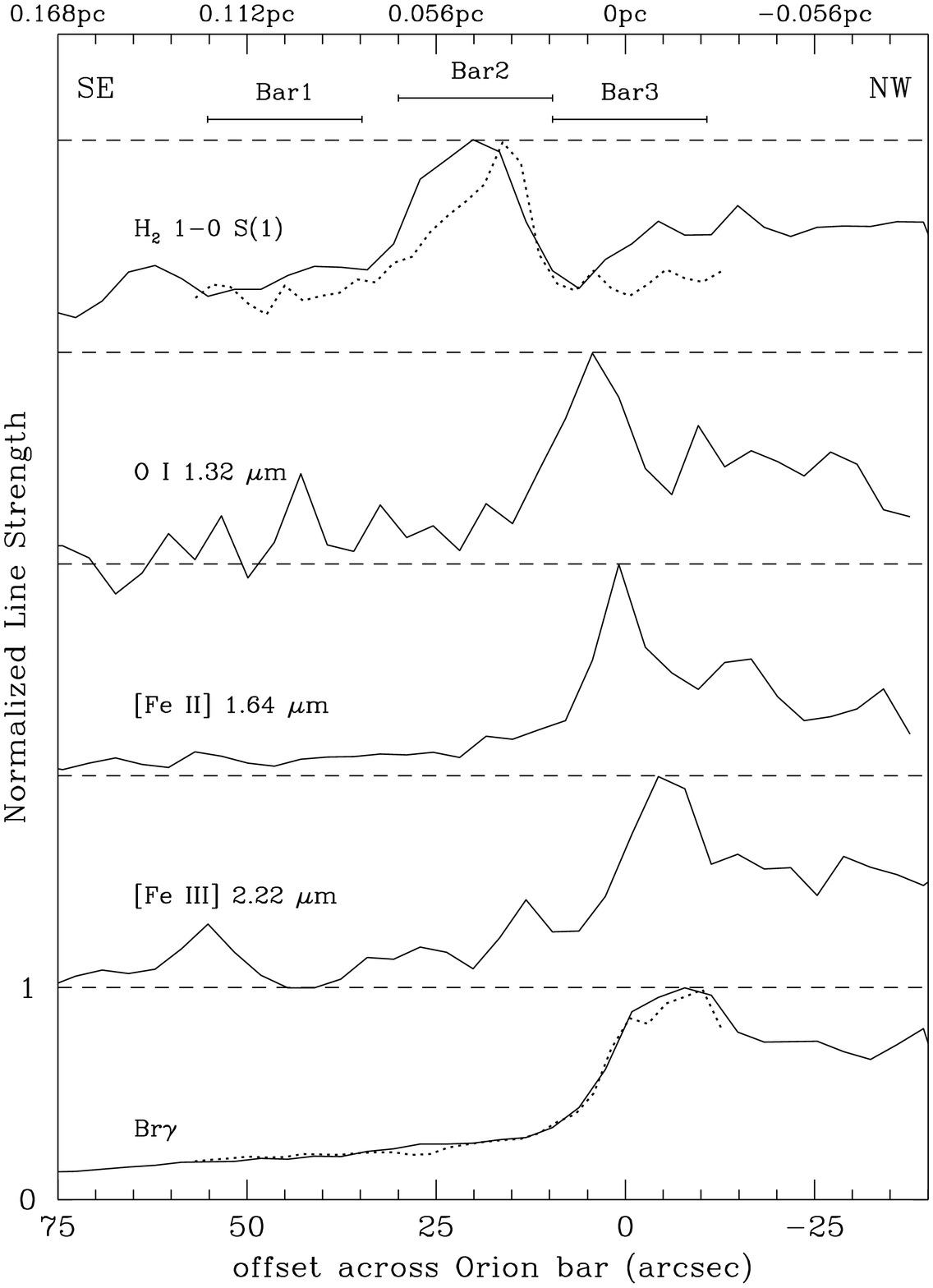}{6.5in}{0}{80}{70}{-245}{0}
\caption{
Spatial variations of several prominent near-IR emission
lines across the Orion bar from SE (left) to NW (right). The solid and dotted
profiles are from the data obtained at the 1.55~m and 2.3~m telescopes with
each resolution element representing $3\farcs5\times3\farcs5$ and
$2\farcs4\times2\farcs4$, respectively. Spectra were extracted from apertures
indicated by the bars at the top of the figure, referred to as Bar1, Bar2, and
Bar3.  The origin corresponds to $\alpha=5^{\rm h}35^{\rm m}19\fs5$,
$\delta=-5\arcdeg24\arcmin53\arcsec$ (2000).
}
\label{fig:probar}
\end{figure}

\begin{figure}
\plotfiddle{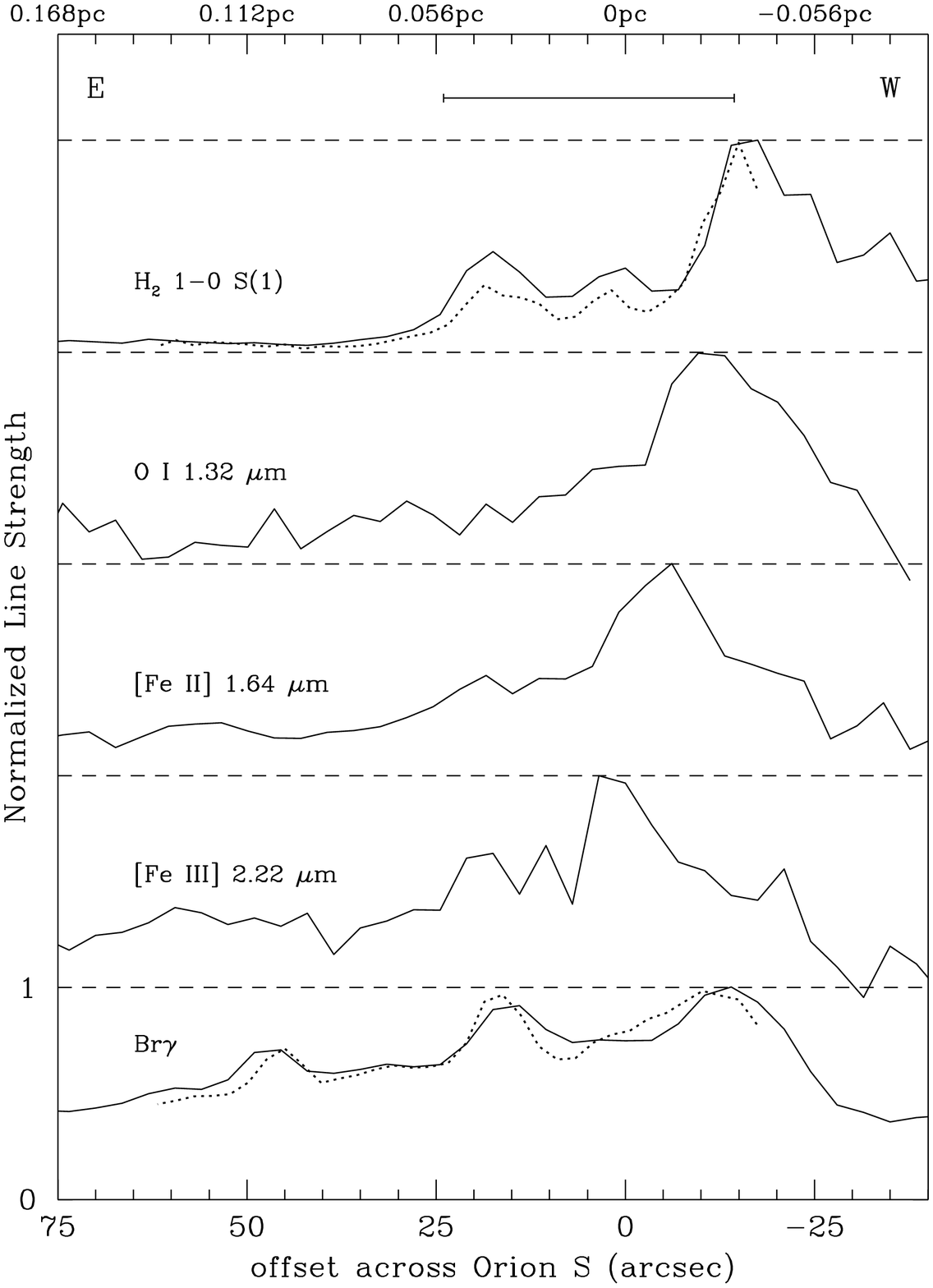}{6.5in}{0}{80}{70}{-245}{0}
\caption{
Spatial variations of several prominent near-IR emission
lines across the Orion~S region from E (left) to W (right). The solid and dotted
profiles are from the data obtained at the 1.55~m and 2.3~m telescopes with
each resolution element representing $3\farcs5\times3\farcs5$ and
$2\farcs4\times2\farcs4$, respectively. A spectrum was extracted from an
aperture indicated by the bar at the top of the figure. The origin corresponds
to $\alpha=5^{\rm h}35^{\rm m}15\fs7$,
$\delta=-5\arcdeg23\arcmin59\arcsec$ (2000).
}
\label{fig:protrap}
\end{figure}

\begin{figure}
\plotfiddle{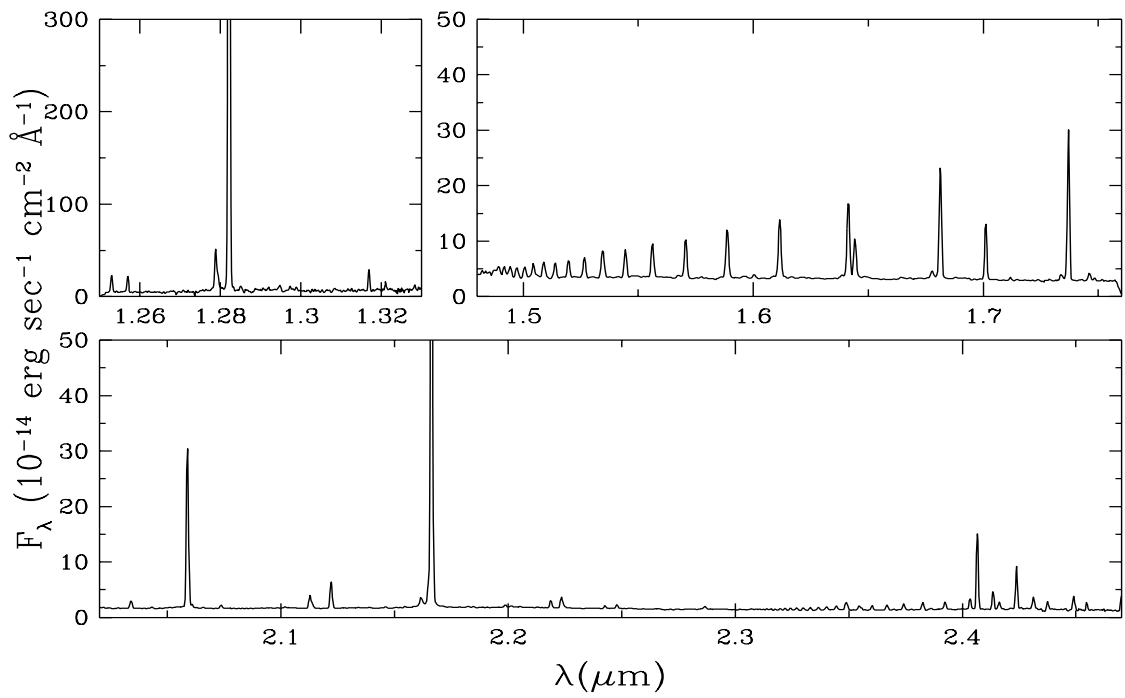}{6.5in}{0}{80}{70}{-245}{0}
\caption{
The near-IR spectrum of Orion~S extracted from the aperture
indicated in Figure~\ref{fig:protrap}.
}
\label{fig:specJHK}
\end{figure}

\begin{figure}
\plotfiddle{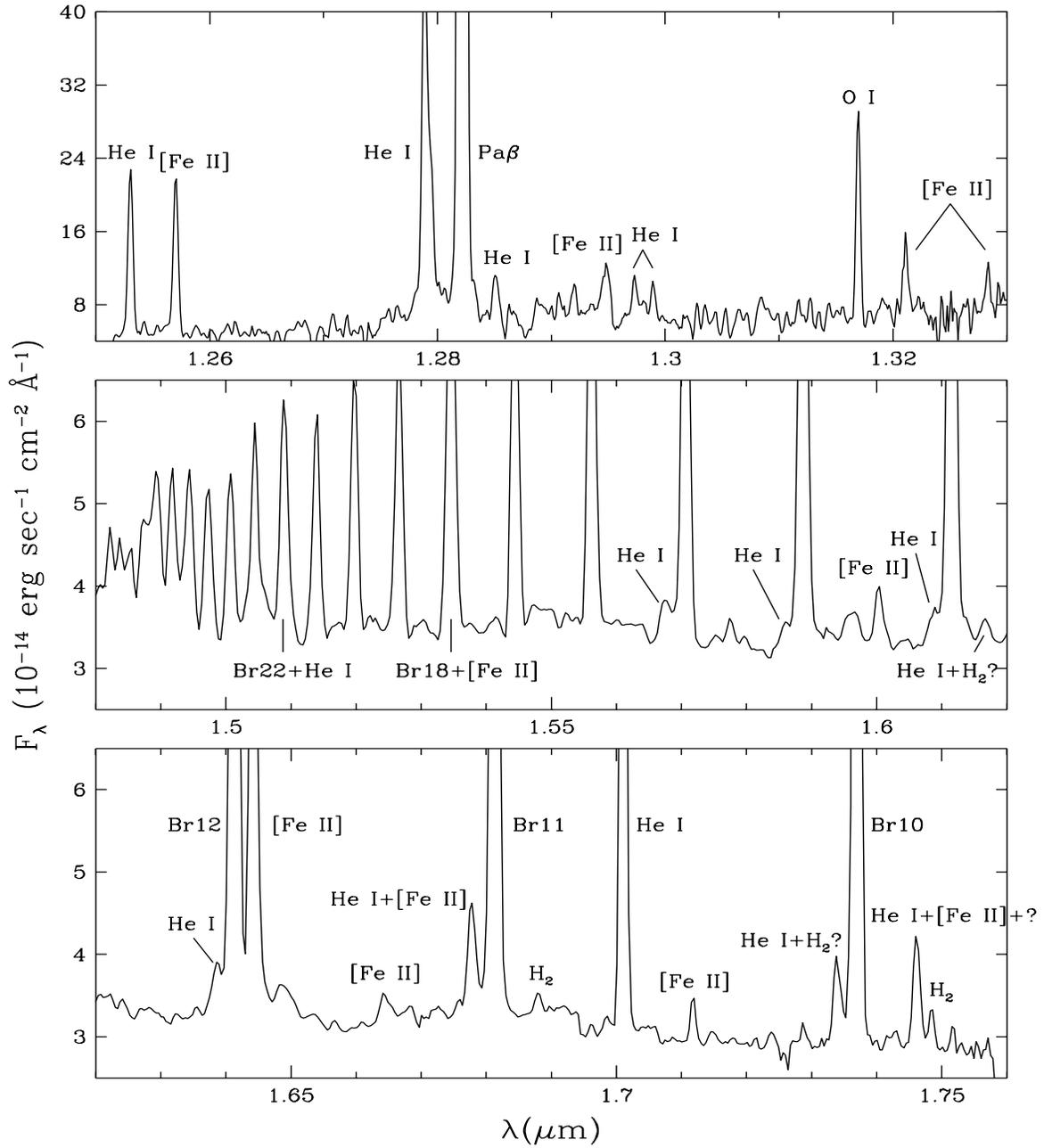}{6.5in}{0}{80}{70}{-245}{0}
\caption{
The spectrum of Orion~S at J and H extracted from the
aperture indicated in Figure~\ref{fig:protrap}. The flux scales for J and H
were derived by using the flux-calibrated K-band spectrum and assuming that
F(Pa$\beta$)/F(Br$\gamma$) and F(Br~lines)/F(Br$\gamma$) match the values
predicted by case B recombination. The unlabeled
emission lines between 1.48 and 1.62~\micron\ are Br28 through Br13.
}
\label{fig:specJH}
\end{figure}

\begin{figure}
\plotfiddle{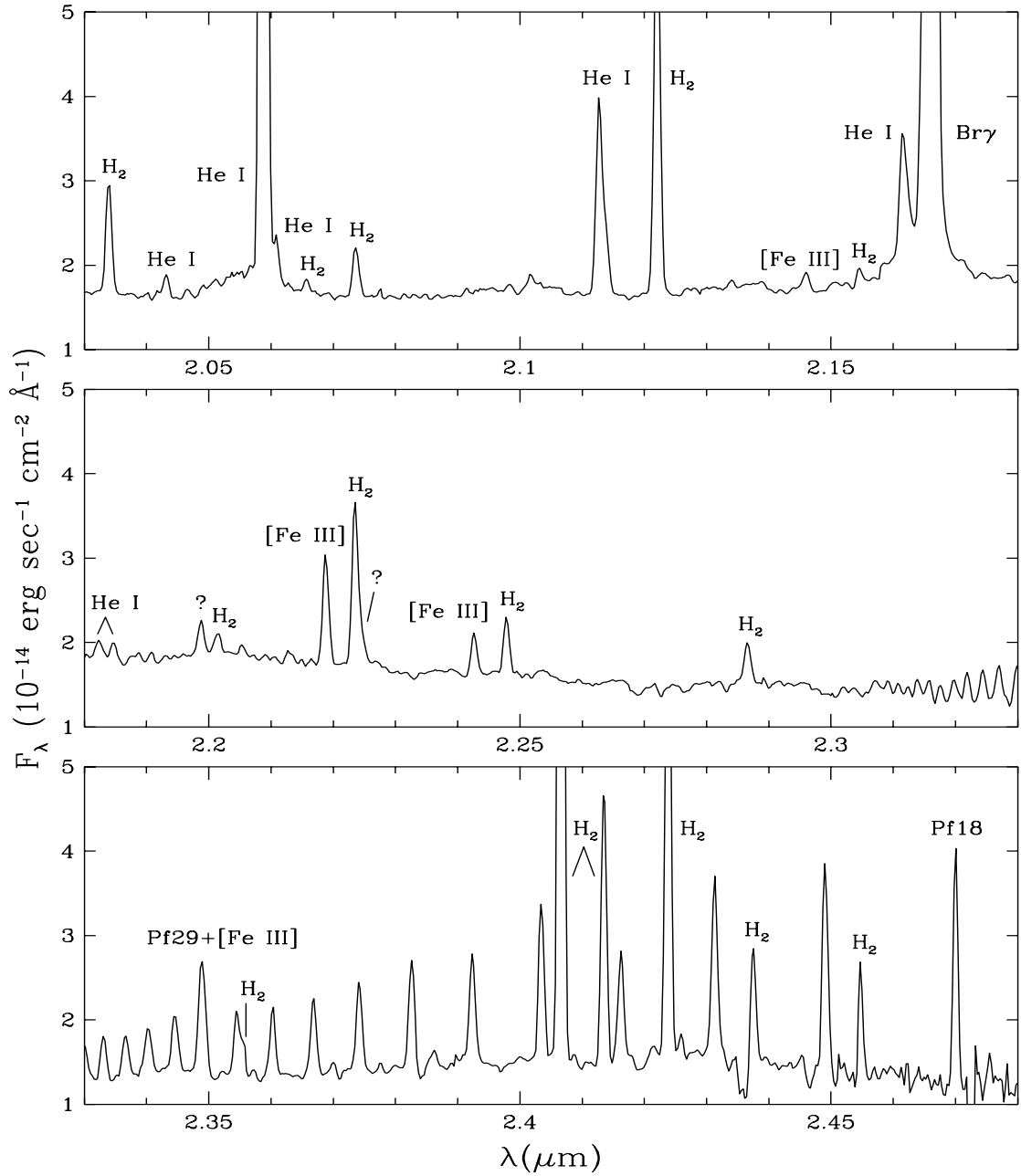}{6.5in}{0}{80}{70}{-245}{0}
\caption{
The spectrum of Orion~S at K extracted from the aperture
indicated in  Figure~\ref{fig:protrap}. The unlabeled emission lines
beyond 2.3~\micron\ are Pf45 through Pf19.
}
\label{fig:specK}
\end{figure}

To select apertures for the extraction of spectra, we examined the spatial
distribution of the major emission lines of each species in the bar and Orion~S,
as illustrated in Figures~\ref{fig:probar} and \ref{fig:protrap}.  Since
the slit was 2 pixels wide and the data were binned by 2 pixels along the slit,
each point in these profiles represents a $2\times2$ pixel region, where one
pixel is $1.75\arcsec$ and $1.2\arcsec$ for the 1.55~m and 2.3~m data,
respectively.  Three $20\arcsec$ apertures shown in Figure~\ref{fig:probar} 
sample three distinct regions of emission in the bar: beyond the bar towards
the molecular cloud (Bar1), on the peak of H$_{2}$ emission (Bar2), and 
in front of the bar towards the H~II region (Bar3).  With the large
spatial overlap of the various emission features in Orion~S, we extracted one
$38\arcsec$ aperture which included the bulk of the emission in each species. 
Due to the poor quality of the 1.55~m data beyond 2.2~\micron\ (see 
\S~\ref{sec:obs}), the K-band spectra from the 2.3~m telescope
were used to analyze emission in [Fe~III] towards Bar3 and Orion~S and H$_{2}$ 
towards Bar1, Bar2, and Orion~S. Since the H$_{2}$ emission towards the H~II
region in Bar3 probably originates from many regions along the line of sight
rather than a well-defined zone of the PDR, we 
do not model the H$_{2}$ in Bar3. The J-, H-, and short K-band (2.0-2.2~\micron)
data were used to examine the line ratios of He~I and [Fe~II] in Bar3 and 
Orion~S.  The spectra for Orion~S are shown in Figures~\ref{fig:specJHK},
\ref{fig:specJH}, and \ref{fig:specK}. The spectra for Bar1, Bar2, and Bar3
are not shown since the combined data appear qualitatively similar to those 
of Orion~S.

Due to guiding errors in observing a standard star with a long-slit 
spectrometer, flux calibration between J, H, and K is relatively 
uncertain ($\pm20$-30\%). To facilitate comparison of line strengths across 
the bands, we calibrated the J- and H-band spectra such that the strengths 
of the hydrogen recombination lines relative to Br$\gamma$ matched the case
B predictions of Storey \& Hummer (1995) for a plasma with 
n$_{\rm e}=10^2$~cm$^{-3}$ and T$_{\rm e}=10^4$~K.  These predicted
line ratios do not depend substantially on density and temperature for the
range of values likely in Orion. This procedure should also correct the
differential extinction (as measured towards the ionized gas) between the
bands.  The Br$\gamma$ fluxes are $4.4\pm0.5$ and 
($15\pm2)\times10^{-12}$~erg~s$^{-1}$~cm$^{-2}$ towards the Bar3 
($3\farcs5\times20\arcsec$) and Orion~S ($3\farcs5\times38\arcsec$)
apertures extracted from the 1.5~m data, respectively. 
In the Bar1 and Bar2 ($2\farcs4\times20\arcsec$) and Orion~S 
($2\farcs4\times38\arcsec$) apertures extracted from the 2.3~m data, the 
fluxes of 1-0~S(1) are $0.8\pm0.15$, $2.0\pm0.4$, and
($5.6\pm1)\times10^{-13}$~erg~s$^{-1}$~cm$^{-2}$, respectively.  
In Figures~\ref{fig:probar} and \ref{fig:protrap}, the maximum values of 
Br$\gamma$ for the bar and Orion~S are 1.04 and 
$1.61\times10^{-12}$~erg~s$^{-1}$~cm$^{-2}$ in $3\farcs5\times3\farcs5$,
corresponding to 3.61 and $5.59\times10^{-3}$~erg~s$^{-1}$~cm$^{-2}$~sr$^{-1}$.
Relative to these Br$\gamma$ measurements, the peak respective values of
H$_{2}$~1-0~S(1), O~I~1.3169~\micron, [Fe~II]~1.6440~\micron, and 
[Fe III]~2.2183~\micron\ are 0.10, 0.12, 0.15, and 0.024 for the bar and 0.13,
0.19, 0.13, and 0.017 for Orion~S.

\subsection{Telluric Correction in Near-IR Emission Line Spectra}
\label{sec:telluric}

As discussed in \S~\ref{sec:obs}, we attempted to remove telluric absorption
in the Orion spectra through division by a standard stellar spectrum.  This
procedure works well with spectra of continuum objects (e.g., stars) but can
be problematic with emission-line spectra. At moderate spectral resolution
(100~km~s$^{-1}$) we cannot observe the true atmospheric transmission, which
is dominated by narrow (10~km~s$^{-1}$), optically thick absorption features. 
As a result, the combination of these telluric absorption features and
intrinsically narrow (5-20~km~s$^{-1}$) astronomical emission lines produces
systematic errors in the measured line fluxes of the latter. For instance, an
emission line passing through a telluric feature is suppressed (or even totally
blocked) and cannot be recovered since the narrow telluric line is smoothed
to moderate resolution. On the other hand, if an emission line is fully
transmitted but falls near a deep absorption feature, division by the
telluric standard spectrum, in which the absorption feature is now smoothed to 
overlap with the emission line, can artificially boost the emission-line flux.

We have modeled these effects with the ultra-high resolution ($\rm R>10^{5}$)
atmospheric transmission spectrum obtained by Livingston \& Wallace (1991).  
For the H$_{2}$ emission region in the bar, we assumed 
V$_{\rm LSR}=10$-11~km~s$^{-1}$ as measured for CO towards the bar by 
Omodaka et al.\ (1994) and van der Werf et al.\ (1996)
and an intrinsic thermal line width of $\rm FWHM=5$~km~s$^{-1}$
($\rm T=1500$~K).  Since the Orion~S region has not been studied extensively,
the representative velocity and line width for this region is uncertain.
CO measurements indicate two lobes of V$_{\rm LSR}=0$-3 and 
11-16~km~s$^{-1}$ (Schmid-Burgk et al.\ 1990; Wilson \& Mauersberger 
1991) which may trace shocked material, while a velocity intermediate
between these and similar to that in the bar would be expected if PDR 
emission is dominant. Consequently, the velocity of the material in the bar
($\sim10$~km~s$^{-1}$) should be roughly appropriate for the Orion~S
region. We then model line widths characteristic of H$_{2}$ from both a 
PDR (5~km~s$^{-1}$) and shocked region (20~km~s$^{-1}$). Finally, to model
the telluric correction for H~I and He~I lines arising in the ionized gas 
we adopt V$_{\rm LSR}\sim$-1~km~s$^{-1}$ (O'Dell 1994) and a line width of
$\rm FWHM=20$~km~s$^{-1}$ (typical for ionized gas in H~II regions).

Figure~\ref{fig:telluric} shows the results of our modeling for all H$_{2}$ 
transitions and the prominent H~I and He~I lines appearing in our 
moderate-resolution K-band spectra of the bar and Orion~S for the range of 
velocities
observable during a year. As expected, comparison of the two models for the 
H$_{2}$ transitions indicates that narrow emission lines are affected more
dramatically.  Lines falling within the telluric CO$_{2}$ bands centered at 
2.01 and 2.06~\micron\ and in the water absorption beyond 2.35~\micron\ are
most susceptible to systematic errors in measured line fluxes.  In fact, 
2-1~S(4) 2.0041~\micron\ and He~2.0587~\micron\ (see \S~\ref{sec:he}) deviate
so rapidly with velocity that they cannot be measured reliably. On the other
hand, lines falling in other regions of the K-band, particularly Br$\gamma$,
1-0~S(1), and 2-1~S(1), are unaffected at the 2\% level over the range of
velocities in question. Although it falls within the water bands, 1-0~Q(4)
2.4375~\micron\ may avoid contamination as long as it is redshifted away
from a broad, opaque atmospheric absorption feature to the immediate blueward
side.  The H$_{2}$ transition 1-0~S(2) 2.0338~\micron\ is also relatively free
of contamination since it falls between the CO$_{2}$ bands. Since these two
transitions, 1-0~Q(4) and 1-0~S(2), share the same upper level and cover a 
large spectral baseline, they can prove useful as an extinction diagnostic
(\S~\ref{sec:h2:ext}).  The other Q-branch transitions show much larger 
deviations in measured line strengths, as do 2-1~S(0) and 3-2~S(1). 
Fortunately, at the time of our observations the error factors were relatively
small ($<25$\%) and do not change significantly with the observed velocity 
(V$_{\rm obs}$) or line width (FWHM).  We find that the prominent H~I, He~I,
and [Fe~II] lines in the J and H
windows are relatively unaffected by problems with telluric correction since
the regions near these lines lack dense groupings of opaque telluric lines. 
However, there are many relatively isolated, narrow ($\sim20$~km~s$^{-1}$)
absorption lines which are opaque and can suppress astronomical 
emission lines which happen to have the appropriate velocity shift, although
this is not a problem for the Orion data presented here.

\section{Discussion}

\subsection{Molecular Hydrogen Lines}
\label{sec:h2}

\subsubsection{Extinction Towards the H$_{2}$ Emission Regions}
\label{sec:h2:ext}

\begin{figure}
\plotfiddle{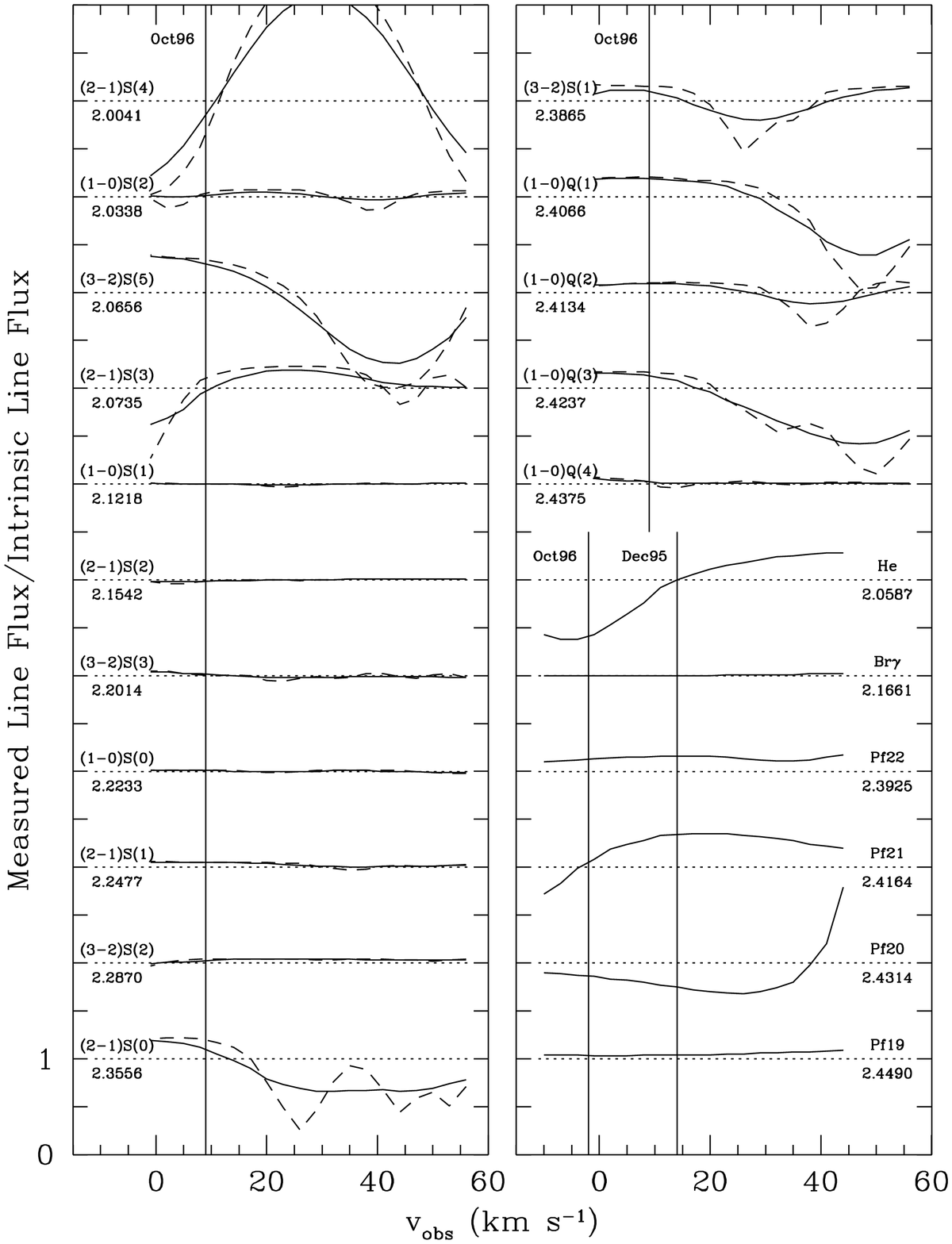}{6.5in}{0}{80}{70}{-245}{0}
\caption{
To illustrate the effect of imperfect telluric correction,
the simulated ratios of measured to intrinsic line fluxes are given for H$_{2}$,
He, and H lines as a function of observed radial velocities of the emission
regions in Orion. An instrumental spectral resolution of $\rm R=3200$ is
assumed with the solid and dashed lines derived with intrinsic FWHMs of 20
and 5~\AA, respectively. Estimates of the velocities of the PDR H$_{2}$, He,
and H regions (taken from the literature) at the dates of our observations
are indicated by the vertical lines.
}
\label{fig:telluric}
\end{figure}

We have detected emission in 17 transitions of H$_{2}$ in our K-band spectra
of the bar and Orion~S.  The H$_{2}$ line ratios measured in the
Bar1, Bar2, and Orion~S apertures (see \S~\ref{sec:extract}) are given
relative to 1-0~S(1) in Table~\ref{tbl:h2}.  The uncertainties in the line
ratios reflect measurement errors in individual lines in addition to 
uncertainties in the relative flux calibration among various grating settings
(\S~\ref{sec:obs}).  As discussed in 
\S~\ref{sec:telluric} and illustrated in Figure~\ref{fig:telluric}, 
imperfect telluric correction produces large systematic errors in 
measurements of several transitions.  Given approximate values of the
velocities and line widths of the emitting regions and the dates of 
observations, we have divided the H$_{2}$ fluxes measured at
2.0656, 2.3556, 2.3865, 2.4066, 2.4134, and 2.4237~\micron\ by 
the estimated correction factors of 1.3, 1.1, 1.1, 1.2, 1.1, and 1.15
(see Figure~\ref{fig:telluric}), respectively. Since emission in 1-0~Q(5) 
2.4547~\micron\ is contaminated substantially by atmospheric absorption, we
do not attempt to correct this line strength and omit it from our discussion.

For comparison to H$_{2}$ spectra
predicted by PDR models (see \S~\ref{sec:h2:compare}), we must deredden
the observed line ratios.  In our K-band data, there are three pairs of 
transitions, 1-0~Q(2)/1-0~S(0), 1-0~Q(3)/1-0~S(1), and 1-0~Q(4)/1-0~S(2), with 
the same upper levels which can be used to estimate the extinction.  Before
using the first of these ratios, we have deblended 1-0~S(0) from an 
unidentified emission line which appears in the red wing of this H$_2$ line at
$\sim2.2245$~\micron.  This unidentified feature appears to arise from an
atomic species (see \S~\ref{sec:other}) and contaminates the 1-0~S(0)
line towards only the Orion~S aperture, which encompasses both molecular and
ionized emission regions.  Using a form of the IR interstellar extinction law
where $\rm A_{\lambda}=A_{K}$($\lambda$/2.2~\micron)$^{-1.75}$, which is 
equivalent to that of Rieke \& Lebofsky (1985), we derived reddenings
of A$_{\rm K}=1.5\pm0.7$, $2.5\pm0.3$, and $1.5\pm0.6$ for Orion~S
and A$_{\rm K}=3.3\pm1$, $2.7\pm0.4$, and $1.3\pm1.7$ for Bar2 from
1-0~Q(2)/1-0~S(0), 1-0~Q(3)/1-0~S(1), and 1-0~Q(4)/1-0~S(2), respectively.
For the Bar1 position, we arrived at A$_{\rm K}=2.8\pm1.3$ and $2.1\pm0.6$
from 1-0~Q(2)/1-0~S(0) and 1-0~Q(3)/1-0~S(1) [1-0~Q(4) was not detected].  The
weighted averages are A$_{\rm K}=2.3\pm0.8$, $2.6\pm0.7$, and $2.1\pm0.5$ for
Bar1, Bar2, and Orion~S.

After applying the atmospheric correction factors to the observed data,
the extinction values were used to deredden the H$_{2}$ line ratios, which
are presented in Table~\ref{tbl:h2} and compared to PDR model predictions
in \S~\ref{sec:h2:compare}.  The origin of these large reddenings towards the
H$_{2}$ emission regions is discussed in \S~\ref{sec:clumps}. We note that
previous observations have also hinted at unexpectedly large extinction.
Sellgren et al.\ (1990) observed Q-branch emission
($\sim2.4$~\micron) which was very strong relative to 1-0~S(1), although 
this effect was attributed at the time to differing spatial scales in the
separate observations of the Q-branch and 1-0~S(1) emission.

\subsubsection{Comparison of the H$_{2}$ Spectra to PDR Model Predictions}
\label{sec:h2:compare}

We now compare the ``corrected" H$_{2}$ line ratios in Table~\ref{tbl:h2} to 
those predicted by recent PDR models of DB96.  In modeling 
the bar and Orion~S, we considered a range of values for 
the density (n$_{\rm H}=10^2$-10$^6$~cm$^{-3}$), temperature 
(T$_{0}=300$-2000~K), and UV field strength ($\chi=10^2$-10$^6$).
Note that the definitions of these parameters, given in DB96, are not 
identical among available PDR models. By
dereddening the data in Table~\ref{tbl:h2}, we have corrected for the effect of 
PDR inclination on the observed line ratios and we need only consider 
face-on geometries. After comparing the various model spectra to the
corrected line ratios for the Bar2 aperture, which includes the peak
of the H$_{2}$ emission, we find an excellent fit with
n$_{\rm H}=10^6$~cm$^{-3}$, $\chi=10^5$, and T$_{0}=1000$~K. As shown in
Table~\ref{tbl:h2}, nearly all line ratios match the data within the 
observational uncertainties. No other combination of parameters produced a 
comparable fit.  These values for the density, temperature, and UV field 
strength agree well with previous estimates towards the dense component of the
PDR (e.g., van der Werf et al.\ 1996), which apparently dominates the
H$_{2}$ emission (see \S~\ref{sec:source}). For the Bar1 aperture,
which includes fainter emission beyond the bar, we find a very good fit with 
n$_{\rm H}=10^6$~cm$^{-3}$, $\chi=10^4$, and T$_{0}=500$~K. 
Marginal fits were produced with 
(n$_{\rm H}=10^5$~cm$^{-3}$, $\chi=10^4$, T$_{0}=1000$~K),
(n$_{\rm H}=10^6$~cm$^{-3}$, $\chi=10^4$, T$_{0}=300$~K), and
(n$_{\rm H}=10^6$~cm$^{-3}$, $\chi=10^5$, T$_{0}=500$, 1000~K).
Consequently, the H$_{2}$ emission in the Bar1 aperture appears to 
originate in a dense PDR, just as in the Bar2 aperture. Towards Bar1
the temperature is lower and the UV field is attenuated, which is consistent
with the location of this emission region behind the bar and deeper
into the molecular cloud.  The data for Orion~S is matched well with
two sets of models which produce nearly identical line ratios,
(n$_{\rm H}=10^5$~cm$^{-3}$, $\chi=10^4$, T$_{0}=1500$~K) and 
(n$_{\rm H}=10^6$~cm$^{-3}$, $\chi=10^5$, T$_{0}=1500$~K). Since
the Orion~S region is closer, at least in projection, to the ionizing stars
than the bar, the model with $\chi=10^5$ is probably more appropriate and
is shown in Table~\ref{tbl:h2}. Therefore, only the temperature differs between
the best models for the Bar2 (T$_{0}=1000$~K) and Orion~S
(T$_{0}=1500$~K) positions.  As seen in Table~\ref{tbl:h2}, the line ratios,
particularly F[2-1~S(1)]/F[1-0~S(1)], are quite distinct between the two 
regions, even with a temperature difference of only 500~K.

For comparison, in Table~\ref{tbl:h2} we also show models of low-density 
fluorescence (n$_{\rm H}=10^4$~cm$^{-3}$, $\chi=10^2$, T$_{0}=500$~K) and 
thermal excitation indicative of a shock ($\rm T=2000$~K, Black \& van Dishoeck 
1987). We have obtained additional K-band spectra towards the BNKL nebula
to compare to the shock model. If we disregard the weak, large-scale emission 
in transitions of $\rm v\geq3$ observed in Orion (Luhman et al.\ 1994),
the dereddened line ratios (A$_{\rm K}=1$) for BNKL match exactly with those
of the thermal model in Table~\ref{tbl:h2}.  However, the corrected line ratios
observed towards the bar and Orion~S do not match the thermal model,
demonstrating that shocks cannot dominate the H$_{2}$ emission in these 
regions. While a dense PDR and a shock produce a similar ratio of
F[2-1~S(1)]/F[1-0~S(1)], the strengths of 1-0~Q(1) and 1-0~Q(2), in 
addition to several weaker transitions shown in Table~\ref{tbl:h2}, relative to
1-0~S(1) clearly discriminate between the two excitation mechanisms.
The low-density model shown in Table~\ref{tbl:h2}, which is insensitive to
changes in $\chi$, also fails to match any of the
observed spectra for the bar and Orion~S. In fact, virtually all of the
H$_2$ emission, even the $\rm v=2$ and 3 transitions, is accounted for
by a dense PDR without the need for a diffuse, pure fluorescent component,
the implications of which are explored in \S~\ref{sec:source}.
The fluorescent transitions at J and H predicted by the favored models in 
Table~\ref{tbl:h2} (under no extinction) fall below our detection limits, 
which were 20\% (J) and 5-10\% (H) of 1-0~S(1) at Bar2 and Orion~S and 30\% (J)
and 20\% (H) of 1-0~S(1) at Bar1.
In low-resolution ($\rm R=950$) K-band long-slit spectra of the bar, Marconi
et al.\ (1997) have recently measured several 1-0 and 2-1 lines with marginal 
detections of a few $\rm v=3$ transitions.  Relying on the ratio of 
2-1~S(1)/1-0~S(1), they estimate a density of 
n$_{\rm H}=3-6\times10^4$~cm$^{-3}$ and thus conclude that there is no evidence
for a substantial filling factor of dense clumps. This is in direct contrast
to our results, where we find excellent agreement between the observed
line ratios of the strongest 16 K-band H$_{2}$ transitions and the predicted
values for a dense PDR (n$_{\rm H}=10^6$~cm$^{-3}$). 

 

Finally, we examine the observed and predicted surface brightness of 
the H$_2$ emission.  For Bar1, Bar2, and Orion~S, we measure 
I[1-0~S(1)$]=0.71\pm0.14$, $1.8\pm0.4$, and
($2.6\pm0.5)\times10^{-4}$~erg~s$^{-1}$~cm$^{-2}$~sr$^{-1}$.
If we correct for the reddening (A$_{\rm K}\sim1$) of the layer between
the ionization front (IF) and H$_2$ emission region (see \S~\ref{sec:clumps}), 
we arrive at I[1-0~S(1)$]=1.8$, 4.5, and 
$6.5\times10^{-4}$~erg~s$^{-1}$~cm$^{-2}$~sr$^{-1}$. The surface brightnesses
predicted by the face-on models in Table~\ref{tbl:h2} are 1.4, 4.7, and 
$12\times10^{-4}$~erg~s$^{-1}$~cm$^{-2}$~sr$^{-1}$ for Bar1, Bar2, and 
Orion~S.  If we include the effect of inclination, the model surface 
brightnesses are up to an order of magnitude greater than the dereddened,
observed values. Since these PDR model predictions only apply to the 
dense component of the PDR (n$_{\rm H}\sim10^6$~cm$^{-3}$), a clump
filling factor less than unity (10-50\%) could bring the observed and
theoretical surface brightnesses into agreement.

\subsubsection{The Source of the H$_{2}$ Spectrum}
\label{sec:source}

In one of the first models of the bar, Hayashi et al.\ (1985) suggested
that the near-IR H$_{2}$ emission originated from shocked material 
($\rm T=2000$~K) along the edge of the expanding H~II region, with low-density
fluorescent emission dominating beyond the bar, deeper into the molecular
cloud.  However, Tielens et al.\ (1993) and van der Werf et al.\ (1996)
concluded that shocks are not an important contributor to the H$_{2}$ emission
since the large shock velocities necessary ($>10$~km s$^{-1}$) are not
observed (although they could be hidden by the inclination of the bar) and
are not expected in an evolved H~II region. Instead, various authors have
proposed a two-component PDR as the source of the H$_{2}$ emission in the bar. 
Tauber et al.\ (1994) and Hogerheijde et al.\ (1995) found clumps on the
scale of $10\arcsec$ and smaller, which they interpreted as dense
regions (n$_{\rm H}\gtrsim10^6$~cm$^{-3}$) of low filling factor within
a diffuse interclump medium (n$_{\rm H}\sim10^4$~cm$^{-3}$). The dense
clumps would then produce the thermal H$_{2}$ spectrum with the low-density
material responsible for the emission in 2-1~S(1), which was assumed to 
be fluorescent. Our measurements of this transition relative to 1-0~S(1)
towards the Bar1 and Bar2 apertures are consistent with those of other
studies (Omodaka et al.\ 1994; Usuda et al.\ 1996; 
van der Werf et al.\ 1996). Most recently, to account for the ratio of
F[2-1~S(1)]/F[1-0~S(1)$]\sim0.33$ behind the bar (equivalent to our Bar1
aperture), van der Werf et al.\ (1996) concluded that a diffuse medium 
(n$_{\rm H}\sim10^4$~cm$^{-3}$) of fluorescent H$_2$ produces the 2-1~S(1)
emission in this region.  However, as demonstrated in \S~\ref{sec:h2:compare},
we find that the entire H$_2$ spectrum, including $\rm v=2$ and 3 transitions,
is reproduced by a dense PDR towards both the Bar1 and Bar2 apertures.
Contrary to previous suggestions, shocked emission at the H$_2$ peak (Bar2)
and low-density fluorescence behind the bar (Bar1) cannot account for
the observed line ratios.  In moving across the bar away from the ionizing
stars (Bar2 to Bar1), it appears that a dense PDR continues to dominate 
the H$_{2}$ emission as the temperature slowly decreases.

Data from far-IR to millimeter wavelengths imply the presence of
a young stellar object $1\farcm5$ south of IRc2, referred to as Orion~S
(McMullen, Mundy, \& Blake 1993, references therein).  While no counterpart
has been detected at shorter wavelengths, prominent emission is apparent
in the immediate vicinity of Orion~S in optical transitions (Pogge, Owen,
\& Atwood 1992), Br$\gamma$, 1-0~S(1), and 2-1~S(1) (Usuda
et al.\ 1996).  In measurements of [S~II] emission, Pogge et al.\ (1992) 
derived significantly higher densities towards Orion~S than in the bar, 
consistent with the difference in densities between the two regions we
find from [Fe~II] (see \S~\ref{sec:fe}). Pogge et al.\ (1992) 
interpreted the Orion~S emission as arising from a dense IF and PDR at
the back of the blister H~II region.  On the other hand, Usuda et al.\ (1996)
found a thermal ratio of F[2-1~S(1)]/F[1-0~S(1)$]=0.1$, similar to what 
we observe, and suggested that the H$_{2}$ emission originates in shocked
material from the unseen, embedded young stellar object. However, as 
demonstrated in \S~\ref{sec:h2:compare}, the overall K-band H$_{2}$ spectrum 
clearly originates in a dense PDR rather than shocked material.
The stratification of the various species in Figure~\ref{fig:protrap}
is also similar to what we observe towards the bar, further supporting 
the inclined PDR nature of the Orion~S emitting region. Compared to the bar, the
spatial distributions of these species imply a smaller inclination and 
more face-on orientation in Orion~S, which is consistent with the somewhat 
lower extinction observed in the H$_{2}$ spectrum of the latter. 

\subsubsection{The Effect of Clumps on PDR Structure}
\label{sec:clumps}

Hogerheijde et al.\ (1995) proposed a model for the bar where the PDR is 
face-on towards the ionized gas, edge-on at the emission 
ridge of the bar, and face-on again southeast of the bar. Furthermore,
Tauber et al.\ (1994) and Hogerheijde et al.\ (1995) found that
clumps do not affect the PDR structure substantially. Although van der 
Werf et al.\ (1996) adopted the geometry proposed by Hogerheijde 
et al.\ (1995), the former concluded that clumps play an important role
in determining the PDR thickness due to the large clump area
filling factor of 50\%, corresponding to an area filling factor of 25\% as 
seen by the Trapezium stars.  This suggestion by van der Werf
et al.\ (1996) may provide a natural explanation for
one of the primary departures of the observed bar structure from that of 
standard PDR models: the anomalous position of the H$_{2}$ emission region
relative to the IF.  In Figure~\ref{fig:probar} we see that
the H$_{2}$ emission reaches a maximum $\sim15\arcsec$ beyond the O~I
emission, which traces the IF.  This fact has been demonstrated previously
through other observations of the H$_{2}$ emission region (Parmar, Lacy,
\& Achtermann 1991) and the IF (van der Werf et al.\ 1996; Omodaka et
al.\ 1994). Nearly coincident with the IF is a ridge of dust emission
features at 3.3~\micron\ (Sellgren et al.\ 1990) and 10~\micron\ (Becklin 
et al.\ 1976).  As discussed in detail by Omodaka et al.\ (1994), in a
homogeneous PDR the UV photons should be absorbed by dust within a thin layer
of A$_{\rm K}\sim0.1$ at the outer edge of the IF where emission in both the
dust and H$_{2}$ is expected to arise, but instead the H$_{2}$ emission occurs
at a position $15\arcsec$ beyond the IF.
Wyrowski et al.\ (1997) have also noted a similar anomaly in observations
of C91$\alpha$ towards the bar.  Whereas standard PDR models (Tielens
\& Hollenbach 1985) predicted the H$_{2}$ emission to arise in a layer between
the IF and C91$\alpha$, Wyrowski et al.\ (1997) found that the H$_{2}$ is 
coincident with the carbon emission.  Both Omodaka et al.\ (1994) and
Wyrowski et al.\ (1997) suggested that the displacement of the H$_{2}$ 
emission peak relative to other species may be explained by either 
shock excitation of the H$_{2}$ or a clumpy PDR in which UV photons
can penetrate the large depth observed between the IF and H$_{2}$ emission
region.

Our observations provide further compelling evidence that the H$_{2}$ emission
arises well beyond the IF. As discussed in \S~\ref{sec:h2:ext}, after
correcting for the extinction towards the ionized gas (and presumably the IF)
and using three pairs of H$_{2}$ transitions with common upper levels, we 
derive extinctions of A$_{\rm K}\sim2.6$ and 2.3 towards the H$_{2}$ peak 
(Bar2) and the weaker H$_{2}$ emission beyond the bar (Bar1).  An internal
extinction of A$_{\rm K}\sim1$ would be apparent in the spectrum if the
PDR is inclined nearly edge-on, but the remaining reddening of 
A$_{\rm K}\sim1$-1.5 must be due to material between the IF and H$_{2}$
emission region.  If we adopt a distance to Orion of $\rm d=450$~pc,
N$_{\rm H}$/A$_{\rm V}\sim2\times10^{21}$~mag$^{-1}$~cm$^{-2}$ (Bohlin, Savage,
\& Drake 1978), and a plausible average density 
(n$_{\rm H}\sim2$-$3\times10^5$~cm$^{-3}$), then the optical depth of this 
$15\arcsec$ layer is consistent with the extinction we measure towards the
H$_{2}$ emission region.  Since we have shown that the 
H$_{2}$ spectrum towards the bar is only reproduced by a dense PDR and not a 
shock (see \S~\ref{sec:h2:compare}), we conclude that a significant amount of UV
radiation is penetrating deep into the atomic gas, probably due to a PDR
with a significant clump filling factor. This is consistent
with the results of van der Werf et al.\ (1996), who found large clumps on the
size of $\sim10\arcsec$. Since this is of order the width of the H$_2$ peak
on the bar, it is not surprising that clumpiness is not obvious in the 
distribution of 1-0~S(1) emission we observe across the bar. 
On the other hand, the dominance of H$_{2}$ fluorescent emission
several arcminutes from the Trapezium stars implies a low clump filling factor
for these distant regions (Luhman et al.\ 1994).  Hence, the low filling factor
($\sim1$\%) assumed by Luhman et al.\ (1997) may be appropriate in their
study of the large-scale H$_{2}$ emission across the Orion~A molecular ridge, 
in contrast to the higher clump filling factor implied by our data in the bar
and Orion~S regions.

\subsection{Helium Lines}
\label{sec:he}

Systematic errors in telluric correction of emission-line spectra are often
overlooked, especially in the previous observations of emission in
He~$\rm 2^1P-2^1S$ 2.0587~\micron.  For instance, we find that the flux in
this line in 1996 October is 60\% of the value measured in 1995 December, as
determined by comparing the ratio of F(He~2.0587~\micron)/F(Br$\gamma$) at each 
date. The model shown in Figure~\ref{fig:telluric} predicts a ratio of 0.43, 
which is fairly close to the measured value considering the sensitivity of this
ratio to the velocities and line widths. A value of 0.60 is produced
by the model by simply increasing the V$_{\rm obs}$ by 3~km~s$^{-1}$ and using
$\rm FWHM=25$~km~s$^{-1}$.  Due to these telluric line effects and the 
complex and uncertain fluorescent excitation of this transition 
(Shields \& DePoy 1994), the $\rm 2^1P-2^1S$ 2.0587~\micron\ emission line 
is not considered further in our analysis. 

In addition to the prominent line at 2.0587~\micron, we have observed
$\sim30$ other He~I transitions in our spectra, most of which have not 
been detected previously. Most of the lines in the $\rm n^{3,1}F-4^{3,1}D$ 
series, appearing at the blue side of the Brackett lines, are quite weak, yet
apparent in our spectra. Note that these H-band transitions were labeled as
unidentified in similar spectra of Hubble 12 (Luhman \& Rieke 1996).  Since the
$\rm 11^{3,1}F-4^{3,1}D$ 1.6780, 1.6786~\micron\ lines are blended with strong 
[Fe~II] emission, these transitions are omitted from the following analysis. 
The $\rm 7^{3,1}L-4^{3,1}F$ 2.1648~\micron\ feature is visible in the wing of
Br$\gamma$, but we cannot measure its strength reliably and omit it from our
discussion as well. In Table~\ref{tbl:he}, we present the fluxes of the
remaining He~I transitions relative to the $\rm 4^3D-3^3P$ 1.7007~\micron\ line
for the Orion~S and Bar3 positions. As an additional test of the He~I models, 
line ratios are given for Hubble~12 (Hora \& Latter 1996; Luhman \& Rieke 1996)
and the ultra-compact H~II region G45.12+0.13 (Lumsden \& Puxley 1996).  We 
compare these line ratios to the predictions of Smits (1996) for collisional 
excitation with T$_{\rm e}=5000$, 10000, and 20000~K at 
n$_{\rm e}=10^{4}$~cm$^{-3}$.  These theoretical line ratios are insensitive
to electron density.  In the least-squares sense, we find that the data are
fit equally well with model predictions at T$_{\rm e}=10000$ and 20000~K.

In the four sources in Table~\ref{tbl:he}, most transitions
have observed line ratios which compare well with predicted values.  
Exceptions include $\rm 4^{3}P-3^{3}S$ 1.2531~\micron\ and
$\rm 7^{3}P-4^{3}S$ 1.7455~\micron\ (after correction for blended H$_{2}$
and [Fe~II]), which are too strong relative to the model predictions for
each source in question.  As discussed by Lumsden \& Puxley (1996)
regarding G45.12+0.13, optically thick UV transitions in $\rm n^{3}P-2^{3}S$
can lead to highly populated $\rm n^{3}P$ levels, which would enhance any
transitions from such levels. Under optically thick conditions, 
Robbins (1968) and Lumsden \& Puxley (1996) calculated
similar enhancement values ranging from $\sim4$ to $\sim2$ for the transitions
with upper levels of 4$^3$P to 7$^3$P, respectively. All detected transitions
from $\rm n^{3}P$ in the compact H~II~regions Hubble~12 
($\rm 4^{3}P-3^{3}S$, $\rm 5^{3}P-3^{3}D$, $\rm 7^{3}P-4^{3}S$, 
$\rm 6^{3}P-4^{3}S$, $\rm 7^{3}P-4^{3}D$) and G45.12+0.13 ($\rm 4^{3}P-3^{3}S$,
$\rm 5^{3}P-3^{3}D$, $\rm 7^{3}P-4^{3}S$) have strengths consistent with 
these predictions. In the more evolved and extended H~II region in Orion,
we find that $\rm 4^{3}P-3^{3}S$ is clearly stronger than model predictions
of Smits (1996), but only by a factor of $\sim$1.5, much less than 
in Hubble~12 and G45.12+0.13.  It is therefore not surprising that 
most of the remaining higher level transitions from $\rm n^{3}P$ 
levels are consistent with little or no enhancement. The only obvious
exception is the transition in $\rm 7^{3}P-4^{3}S$ 1.7455~\micron, 
which is very strong compared to model predictions of Smits (1996) even after 
subtraction of the expected contribution of [Fe~II] 1.7454~\micron.  This
feature may include a third unidentified line blended with He~I and [Fe~II],
possibly C~I at 1.7453~\micron\ (Outred 1978). We do not expect fluorescent
emission in H$_2$ 7-5~Q(3) at 1.7463~\micron\ to contribute significantly to
this blend in the Orion bar or Orion~S. In conclusion, it appears
that the transitions in $\rm n^{3}P-2^{3}S$ are optically thick in 
Hubble~12 and G45.12+0.13, while the effect is less extreme in Orion. 
This opacity effect may explain the anomalous decrement in these UV transitions
reported by Martin et al.\ (1996) in observations of Orion.

\subsection{Iron Lines}
\label{sec:fe}

We have detected numerous near-IR [Fe~II] and [Fe~III] lines in the Bar3
and Orion~S apertures, which are given relative to the lines at 
1.6440~\micron\ and 2.2183~\micron\ in Tables~\ref{tbl:feii} and 
\ref{tbl:feiii}.  As discussed in \S~\ref{sec:he}, enhancement in the He~I
transition at 1.7455~\micron\ and the presence of a blended unidentified line 
are likely, so we cannot reliably measure the flux of [Fe~II]~1.7454~\micron.
Since several pairs of [Fe~II] transitions possess the same upper levels, 
their relative strengths are determined by the A-values and line wavelengths 
alone, making them useful for reddening estimates. In practice, since most of
these lines are fairly weak and cover short wavelength baselines, the
resulting extinction measurements are highly uncertain. However, the two
strongest near-IR [Fe~II] transitions are substantially separated in wavelength
and have the same upper level, with an expected ratio of
F(1.2570~\micron)/F(1.6440~\micron$)=1.36$.  Since all line ratios are 
corrected for the reddening towards the ionized hydrogen through the 
flux calibration in \S~\ref{sec:extract}, we therefore derive 
A$_{\rm K}\sim0.3\pm0.3$ as the differential line-of-sight extinction between 
the regions of H~II and Fe~II in both the bar and Orion~S.

In recent years, there has been some question as to the physical conditions,
particularly the density, in the Orion [Fe~II] emission regions.
Bautista, Pradhan, \& Osterbrock (1994) and Bautista \& Pradhan (1997)
examined optical and J-band 
[Fe~II] lines towards Orion, concluding that the emission in [Fe~II]
is produced from high-density gas (n$_{\rm e}\sim2\times10^6$~cm$^{-3}$).
With improved spectroscopic data
and new model calculations, Baldwin et al.\ (1996) found that the optical 
[Fe~II] line ratios indicated highly-populated upper levels in Fe~II, in
agreement with observations of Bautista et al.\ (1994). But the latter
authors attributed such level populations to collisional excitation in a dense 
environment, whereas Baldwin et al.\ (1996) proposed that UV pumping can
also effectively populate the upper levels of Fe~II in a low-density gas 
(n$_{\rm e}\sim10^4$~cm$^{-3}$).

Since the IR transitions in [Fe~II] occur between lower levels, 
these line ratios are insensitive to continuum pumping and depend only on
collisional excitation. Therefore, the IR [Fe~II] lines represent an
excellent density diagnostic.  For instance, Baldwin et al.\ (1996) noted that
the strength of the line at 1.2570~\micron\ (Lowe, Moorehead, \& Wehlau 1979) 
relative to the optical lines implies low densities. Marconi et al.\ (1997) 
have also used the strength of one [Fe~II] line relative to the transition at
1.6440~\micron, which was completely blended with Br12 at their resolution
($\rm R=700$), to estimate a density of n$_{\rm e}\sim10^4$~cm$^{-3}$ towards 
the bar. However, with the large number of J- and H-band [Fe~II] lines that
we have observed towards the bar and Orion~S, we can perform a much more
systematic measurement of the density, similar to that done for Hubble~12 and
the starburst galaxy NGC~253 (Luhman \& Rieke 1996; Engelbracht et al.\ 1997). 
In Table~\ref{tbl:feii} we compare
the observed line ratios with those predicted by the collisional model
predictions of Bautista \& Pradhan (1996) for n$_{\rm e}=10^3$, 10$^4$,
and 10$^6$~cm$^{-3}$ at T$_{\rm e}=10^4$~K.  The J-band transitions 
are consistent within the uncertainties with all three models, while
the H-band lines indicate densities of n$_{\rm e}\sim10^4$~cm$^{-3}$
for Orion~S and n$_{\rm e}\sim10^3$-10$^4$~cm$^{-3}$ for the bar. 
K-band emission in [Fe~III] is also strong towards Orion, as noted by
DePoy \& Pogge (1994) in their observations of the Orion~S region. 
We have measured four near-IR [Fe~III] transitions and
compared the line ratios to the model predictions of Keenan et al.\ (1992).
As shown in Table~\ref{tbl:feiii}, the line ratios are all roughly consistent
with the values predicted for n$_{\rm e}\sim10^4$~cm$^{-3}$.
The differing density measurements derived in the optical
and IR can be explained in a model where the near-IR emission arises
in low-density, fully-ionized material while the optical transitions are
produced predominantly in high-density, partially-ionized regions,
as proposed by Bautista \& Pradhan (1997). 

\subsection{Other Lines}
\label{sec:other}

There are three moderately strong unidentified emission lines in our spectra
of the bar and Orion~S. A feature appears at 2.1987~\micron\ which has
been observed in the spectra of Hubble~12 and other planetary nebulae (Luhman
\& Rieke 1996; Hora \& Latter 1996). A second unidentified line, which 
apparently has never been detected before, is visible at 2.2245~\micron\ and
is partially blended with 1-0~S(0) 2.2233~\micron\ in the spectrum of Orion~S
in Figure~\ref{fig:specK}. The line does not appear towards Bar2 but is 
clearly resolved from weak 1-0~S(0) 2.2233~\micron\ emission in the Bar3
spectrum. Consequently, both of these unidentified emission lines appear to 
arise from atomic species since they are only detected towards the H~II regions
in Orion and Hubble~12.  In preparation for the observations presented here,
we obtained low-resolution ($\rm R=1200$) spectra of the bar and Orion~S
at J, H, and K.  In the J-band spectrum, which extended to shorter
wavelengths than the moderate-resolution spectrum, we detect the He~I transition
at 1.1973~\micron\ and an unidentified line at $1.1889\pm0.0002$~\micron.
Possible line identifications include C~I (1.1896~\micron) and Si~I
(1.188~\micron).  An emission line at a similar wavelength is also detected
in the starburst galaxies NGC~253 (Engelbracht et al.\ 1997) and NGC~6240
(Engelbracht, Rieke, \& Rieke 1998; Kulesa et al.\ 1998).

\section{Conclusion}

We have presented near-IR spectra ($\rm R\sim3000$) which provide new
insight into the physical conditions and excitation mechanisms present
in the Orion bar and Orion~S regions:

\begin{enumerate}

\item
After comparing the relative strengths of 16 H$_{2}$ lines to those produced
by models of dense PDRs (Draine \& Bertoldi 1996) and thermal excitation, we 
rule out shocks as a significant contributor to emission in either the bar
or Orion~S. While similar ratios of F[2-1~S(1)]/F[1-0~S(1)] arise in dense PDRs
and shocks, we demonstrate that several other H$_{2}$ lines in our data are 
powerful diagnostics in distinguishing between the two excitation mechanisms.
Towards the peak of H$_{2}$ emission on the bar, the line ratios
are in excellent agreement with PDR model predictions 
(n$_{\rm H}=10^6$~cm$^{-3}$, $\chi=10^5$, T$_{0}=1000$~K). In
the region behind the bar and away from the ionizing stars, the entire
H$_2$ spectrum, including the transitions with $\rm v=2$ and 3, implies 
the same density regime but with a lower temperature (T$_{0}=500$~K)
and UV field strength ($\chi=10^4$). This is in contrast to previous
suggestions that low-density fluorescing material is required to produce
the observed 2-1~S(1) emission.  In the Orion~S region, the H$_2$ line
ratios reveal the presence of a dense PDR similar to that of the bar, but
with a higher temperature (T$_{0}=1500$~K) in Orion~S. The
spatial stratification of several species (O~I, H~I, [Fe~II], [Fe~III], 
H$_{2}$) is similar between the bar and Orion~S, indicating an inclined
PDR in each region. The geometrical enhancement of line emission
observed in these two sources may be a common phenomenon in other bright
regions of the ISM as well.

\item
Using three pairs of H$_{2}$ lines with common upper levels, we derive 
the extinction towards the H$_{2}$ emission regions in the bar
(A$_{\rm K}=2.6\pm0.7$) and Orion~S (A$_{\rm K}=2.1\pm0.5$). Since
an edge-on PDR should exhibit a reddening of only A$_{\rm K}\sim1$, the 
remaining extinction is apparently due to a thick layer (A$_{\rm K}=1$-1.5) of
material between the ionization front and the H$_2$ emitting gas.  While
this layer is predicted to be rather thin (A$_{\rm K}\sim0.1$) in homogeneous 
PDR models, a significant filling factor of dense clumps 
(n$_{\rm H}\sim10^6$~cm$^{-3}$) could allow UV photons to penetrate further
into the PDR and produce the H$_2$ peak at the large depth we observe 
in the bar and Orion~S. 

\item
Most of the $\sim30$ He~I emission lines have relative strengths 
consistent with models of collisional excitation (Smits 1996).
A few transitions, particularly $\rm 4^3P-3^3S$ 1.2531~\micron, appear 
stronger than predicted, possibly due to opacity in the UV transitions
of $\rm n^{3}P-2^{3}S$. After comparing the He~I line ratios we observed in
Orion to those measured in Hubble~12 and G45.12+0.13 (Hora \& Latter 1996; 
Luhman \& Rieke 1996; Lumsden \& Puxley 1996), we find that this effect
is more pronounced in the latter two sources.

\item
The line ratios of 10 near-IR [Fe~II] transitions agree well with the values 
predicted by Bautista \& Pradhan (1996) for collisional excitation towards the
bar (n$_{\rm e}\sim10^3$-10$^4$~cm$^{-3}$) and Orion~S
(n$_{\rm e}\sim10^4$~cm$^{-3}$).  In addition, the relative
strengths of four [Fe~III] transitions match predictions by Keenan et 
al.\ (1992) for a density of n$_{\rm e}\sim10^4$~cm$^{-3}$.
When these results are combined with measurements of higher densities
(n$_{\rm e}\sim10^6$~cm$^{-3}$) from optical [Fe~II] data,
it appears that dense, partially-ionized
regions are responsible for the optical emission while fully-ionized gas
dominates the near-IR [Fe~II] transitions, which is consistent with recent
model predictions of Bautista \& Pradhan (1997). 

\end{enumerate}

\acknowledgements
We thank G. Rieke and M. Rieke for help in obtaining the data. 
We are grateful to M. Bautista, F. Bertoldi, B. Draine, and D. Smits
for useful advice and access to their model calculations. We thank
P. Martin for communicating results prior to publication and G. Rieke
for comments on the manuscript.  K. L. acknowledges support from NASA 
grant NAGW-4083 under the Origins of Solar Systems program.  C. E.
and M. L. were supported by NSF grant AST95-29190 and the Office of 
Naval Research, respectively.

\newpage

\begin{table}
\dummytable\label{tbl:h2}
\end{table}

\clearpage

\begin{deluxetable}{llllllllll}
\scriptsize
\tablewidth{0pt}
\tablecaption{He Line Ratios\tablenotemark{a} \label{tbl:he}}
\tablehead{\colhead{} & \colhead{} & \multicolumn{4}{c}{Observed} &
\colhead{} & \multicolumn{3}{c}{Model\tablenotemark{b}} \\
\cline{3-6}\cline{8-10}
\colhead{$\lambda_{\rm vac}$(\micron)\tablenotemark{c}} & \colhead{transition} &
\colhead{Bar3} & \colhead{Orion~S} & \colhead{Hubble~12\tablenotemark{d}} & 
\colhead{G45.12+0.13\tablenotemark{e}} &
\colhead{} & \colhead{T$_3=5$} & \colhead{T$_3=10$} & \colhead{T$_3=20$}}
\startdata
1.2531 & $\rm 4^3P-3^3S$ & $0.64\pm0.10$ & $0.87\pm0.09$ & $1.42\pm0.10$ & $1.38\pm0.08$ & & $0.36$ & $0.44$ & $0.56$ \nl
1.2789,93 & $\rm 5^{3,1}F-3^{3,1}D$ & $2.44\pm0.15$ & $2.61\pm0.20$ & $2.37\pm$? & $5.40\pm2.00$ & & $3.49$ & $3.06$ & $2.67$ \nl
1.2850 & $\rm 5^3S-3^3P$ & $0.11\pm0.11$ & $0.31\pm0.11$ & \nodata & \nodata & & $0.08$ & $0.12$ & $0.18$ \nl
1.2972 & $\rm 5^1D-3^1P$ & $0.11\pm0.11$ & $0.13\pm0.10$ & $0.52\pm0.05$\tablenotemark{f} & $0.52\pm0.05$\tablenotemark{f} & & $0.23$ & $0.23$ & $0.24$ \nl
1.2989 & $\rm 5^3P-3^3D$ & $0.17\pm0.11$ & $0.14\pm0.10$ & \nodata & \nodata & & $0.10$ & $0.13$ & $0.16$ \nl
1.5088\tablenotemark{g} & $\rm 4^1P-3^1S$ & $0.05\pm0.06$ & $0.09\pm0.04$ & $0.23\pm0.03$ & $0.19\pm0.05$ & & $0.15$ & $0.18$ & $0.21$ \nl
1.5678,82 & $\rm 15^{3,1}F-4^{3,1}D$ & $0.02\pm0.02$ & $0.05\pm0.02$ & $0.02\pm0.02$ & \nodata & & $0.02$ & $0.02$ & $0.02$ \nl
1.5857,62 & $\rm 14^{3,1}F-4^{3,1}D$ & $0.01\pm0.02$ & $0.03\pm0.02$ & $0.02\pm0.02$ & \nodata & & $0.04$ & $0.04$ & $0.03$ \nl
1.6085,90 & $\rm 13^{3,1}F-4^{3,1}D$ & $0.02\pm0.02$ & $0.04\pm0.02$ & $0.02\pm0.02$\tablenotemark{i} & \nodata & & $0.05$ & $0.04$ & $0.04$ \nl
1.6165 & $\rm 11^{3}D-4^{3}P$ & $0.02\pm0.02$ & $0.03\pm0.02$ & \nodata & \nodata & & $0.02$ & $0.02$ & $0.02$ \nl
1.6382,88 & $\rm 12^{3,1}F-4^{3,1}D$ & $0.04\pm0.02$ & $0.06\pm0.02$ & $0.05\pm0.01$ & \nodata & & $0.06$ & $0.06$ & $0.05$ \nl
1.6678 & $\rm 10^3D-4^3P$ & $0.01\pm0.01$ & $0.01\pm0.02$ & $0.04\pm0.01$ & \nodata & & $0.02$ & $0.02$ & $0.02$ \nl
1.7007 & $\rm 4^3D-3^3P$ & 1.00 & 1.00 & 1.00 & \nodata & & 1.00 & 1.00 & 1.00 \nl
1.7334,40 & $\rm 10^{3,1}F-4^{3,1}D$ & $0.15\pm0.04$ & $0.11\pm0.02$ & $0.08\pm0.03$\tablenotemark{i} & \nodata & & $0.10$ & $0.09$ & $0.09$ \nl
1.7428 & $\rm 9^3D-4^3P$ & $0.04\pm0.02$ & $0.02\pm0.01$ & $0.02\pm0.02$ & \nodata & & $0.03$ & $0.03$ & $0.03$ \nl
1.7455\tablenotemark{h} & $\rm 7^3P-4^3S$ & $0.15\pm0.05$ & $0.11\pm0.04$ & $0.14\pm0.02$\tablenotemark{i} & $0.10\pm0.05$ & & $0.02$ & $0.03$ & $0.03$ \nl
2.0430 & $\rm 6^3P-4^3S$ & $0.03\pm0.02$ & $0.02\pm0.03$ & $0.06\pm0.01$ & \nodata & & $0.03$ & $0.03$ & $0.04$ \nl
2.0587 & $\rm 2^1P-2^1S$ & $8.08\pm0.33$ & $5.13\pm0.16$ & $4.65\pm0.11$ & $11.00\pm0.70$ & & \nodata & \nodata & \nodata \nl
2.1127,38 & $\rm 4^{3,1}S-3^{3,1}P$ & $0.40\pm0.04$ & $0.34\pm0.03$ & $0.53\pm0.02$ & $0.48\pm0.05$ & & $0.25$ & $0.34$ & $0.49$ \nl
2.1613,22 & $\rm 7^{3,1}F-4^{3,1}D$ & $0.23\pm0.04$ & $0.25\pm0.03$ & $0.27\pm0.02$ & \nodata & & $0.29$ & $0.27$ & $0.24$ \nl
2.1821 & $\rm 7^3P-4^3D$ & $0.02\pm0.02$ & $0.03\pm0.01$ & $0.04\pm0.02$ & \nodata & & $0.02$ & $0.02$ & $0.03$ \nl
2.1846 & $\rm 7^1D-4^1P$ & $0.01\pm0.02$ & $0.02\pm0.01$ & $0.02\pm0.02$ & \nodata & & $0.02$ & $0.02$ & $0.02$ \nl
\enddata
\tablenotetext{a}{Line ratios are relative to $\rm 4^3D-3^3P$ 1.7007~\micron.
For Bar3 and Orion~S, F(He~1.7007~\micron)/F(Br$\gamma)=0.116\pm0.005$ and
$0.112\pm0.004$ where
F(Br$\gamma)=(4.4\pm0.5)\times10^{-12}$~erg~s$^{-1}$~cm$^{-2}$ and
$(1.5\pm0.2)\times10^{-11}$~erg~s$^{-1}$~cm$^{-2}$.}
\tablenotetext{b}{Smits (1996) with n$_{\rm e}=10^4$~cm$^{-3}$. Model values 
for the transition at 2.0587~\micron\ are highly uncertain.}
\tablenotetext{c}{Martin (1973).}
\tablenotetext{d}{J-band data from Hora \& Latter (1996). H- and K-band data 
from Luhman \& Rieke (1996).}
\tablenotetext{e}{Lumsden \& Puxley (1996).}
\tablenotetext{f}{Blended with the transition at 1.2989~\micron.}
\tablenotetext{g}{The flux in this line was estimated by assuming the 
contribution of Br22 is that derived from the measured Br21 flux and the ratio 
of Br21 to Br22 predicted by case B recombination with 
n$_{\rm e}=10^2$~cm$^{-3}$ and T$_{\rm e}=10^4$~K (Storey \& Hummer 1995).}
\tablenotetext{h}{Blended [Fe~II] emission was subtracted by 
using the theoretical line strength relative to [Fe~II] 1.6440~\micron\
(Bautista \& Pradhan 1996).}
\tablenotetext{i}{Blended fluorescent H$_{2}$ emission was subtracted by
using the theoretical pure fluorescent line strength relative to 1-0~S(1) 
2.1220~\micron\ (Black \& van Dishoeck 1987).}
\end{deluxetable}

\clearpage

\begin{deluxetable}{llllllll}
\tablewidth{0pt}
\tablecaption{[\ion{Fe}{2}] Line Ratios\tablenotemark{a} \label{tbl:feii}}
\tablehead{
\colhead{} & \colhead{} & \multicolumn{2}{c}{Observed} & \colhead{} & 
\multicolumn{3}{c}{Model\tablenotemark{c}}\\
\cline{3-4}\cline{6-8}
\colhead{$\lambda_{\rm vac}$\tablenotemark{b}} & \colhead{transition} & 
\colhead{Bar3} & \colhead{Orion~S} & \colhead{} & \colhead{n$_{\rm e}$=10$^3$}
& \colhead{n$_{\rm e}$=10$^4$} & \colhead{n$_{\rm e}$=10$^6$}}
\startdata
1.2570 & $\rm ^4D_{7/2}-\ ^6D_{9/2}$ & $1.10\pm0.11$ & $1.03\pm0.11$ & & 1.36 &
1.36 & 1.36 \nl
1.2946 & $\rm ^4D_{5/2}-\ ^6D_{5/2}$ & $0.16\pm0.12$ & $0.36\pm0.17$bl? & & 0.06
& 0.19 & 0.36 \nl
1.3209 & $\rm ^4D_{7/2}-\ ^6D_{7/2}$ & $0.41\pm0.12$ & $0.46\pm0.13$ & & 0.36 &
0.36 & 0.36 \nl
1.3281 & $\rm ^4D_{5/2}-\ ^6D_{3/2}$ & $<0.14$ & $0.21\pm0.11$ & & 0.04 & 0.12 &
0.22 \nl
1.5339\tablenotemark{d} & $\rm ^4D_{5/2}-\ ^4F_{9/2}$ & $0.13\pm0.05$ & 
$0.23\pm0.05$ & & 0.07 & 0.20 & 0.38 \nl
1.5999 & $\rm ^4D_{3/2}-\ ^4F_{7/2}$ & $0.03\pm0.03$ & $0.08\pm0.03$ & & 0.03 &
0.14 & 0.31 \nl
1.6440 & $\rm ^4D_{7/2}-\ ^4F_{9/2}$ & $1.00$ & $1.00$ & & 1.00 & 1.00 & 
1.00 \nl
1.6642 & $\rm ^4D_{1/2}-\ ^4F_{5/2}$ & $0.03\pm0.03$ & $0.05\pm0.04$ & & 0.02
& 0.08 & 0.16 \nl
1.6773\tablenotemark{e} & $\rm ^4D_{5/2}-\ ^4F_{7/2}$ & $0.08\pm0.04$ & 
$0.10\pm0.03$ & & 0.05 & 0.15 & 0.28 \nl
1.7116 & $\rm ^4D_{3/2}-\ ^4F_{5/2}$ & $0.03\pm0.02$ & $0.05\pm0.01$ & & 0.01 &
0.04 & 0.08 \nl
\enddata
\tablenotetext{a}{Line ratios are relative to $\rm ^4D_{7/2}-\ ^4F_{9/2}$
1.6440~\micron.  For Bar3 and Orion~S,
F([Fe~II]~1.6440~\micron)/F(Br$\gamma)=0.109\pm0.006$ and $0.090\pm0.004$ where
F(Br$\gamma)=(4.4\pm0.5)\times10^{-12}$~erg~s$^{-1}$~cm$^{-2}$ and
$(1.5\pm0.2)\times10^{-11}$~erg~s$^{-1}$~cm$^{-2}$.}
\tablenotetext{b}{Johannson (1978).}
\tablenotetext{c}{Collisional excitation with T$_{\rm e}=10^4$~K (Bautista 
\& Pradhan 1996).}
\tablenotetext{d}{Br18 was subtracted by interpolating between Br19 and Br17.}
\tablenotetext{e}{He $\rm 11^{3,1}F-4^{3,1}D$ was subtracted by interpolating
between He $\rm 12^{3,1}F-4^{3,1}D$ and $\rm 10^{3,1}F-4^{3,1}D$.}
\end{deluxetable}

\clearpage

\begin{deluxetable}{llllllll}
\tablewidth{0pt}
\tablecaption{[\ion{Fe}{3}] Line Ratios\tablenotemark{a} \label{tbl:feiii}}
\tablehead{\colhead{} & \colhead{} & \multicolumn{2}{c}{Observed} & \colhead{}
& \multicolumn{3}{c}{Model\tablenotemark{c}} \\
\cline{3-4} \cline{6-8}
\colhead{$\lambda_{\rm vac}$\tablenotemark{b}} & \colhead{transition} &
\colhead{Orion~S} & \colhead{Bar3} & \colhead{} &\colhead{n$_{\rm e}$=10$^3$}
& \colhead{n$_{\rm e}$=10$^4$} & \colhead{n$_{\rm e}$=10$^5$}}
\startdata
2.1457 & $\rm ^3G_3-\ ^3H_4$ & $0.11\pm0.11$ & $0.17\pm0.06$ & & 0.10 &
0.17 & 0.34 \nl
2.2183 & $\rm ^3G_5-\ ^3H_6$ & $1.00$ & $1.00$ & & 1.00 & 1.00 & 1.00 \nl
2.2427 & $\rm ^3G_4-\ ^3H_4$ & $0.34\pm0.12$ & $0.38\pm0.07$ & & 0.26 &
0.29 & 0.38 \nl
2.3485\tablenotemark{d} & $\rm ^3G_5-\ ^3H_5$ & $0.89\pm0.17$ & $0.77\pm0.09$ 
& & 0.66 & 0.66 & 0.66 \nl
\enddata
\tablenotetext{a}{Line ratios are relative to $\rm ^3G_5-\ ^3H_6$ 
2.2183~\micron.  For Bar3 and Orion~S, 
F([Fe~III]~2.2183~\micron)/F(Br$\gamma)=0.015\pm0.001$ and $0.016\pm0.001$ where
F(Br$\gamma)=(4.4\pm0.5)\times10^{-12}$~erg~s$^{-1}$~cm$^{-2}$ and
$(1.5\pm0.2)\times10^{-11}$~erg~s$^{-1}$~cm$^{-2}$.}
\tablenotetext{b}{Sugar \& Corliss (1985).}
\tablenotetext{c}{Collisional excitation with T$_{\rm e}=10^4$~K (Keenan
et al.\ 1992).}
\tablenotetext{d}{The flux in this line was estimated by assuming the
contribution of Pf29 is that derived from the measured Br$\gamma$ flux and the
ratio of Pf29 to Br$\gamma$ predicted by case B recombination with 
n$_{\rm e}=10^2$~cm$^{-3}$ and T$_{\rm e}=10^4$~K (Storey \& Hummer 1995).}
\end{deluxetable}

\clearpage
\pagestyle{empty}
 
\clearpage

\clearpage

\clearpage

\clearpage

\clearpage

\clearpage


\begin{references}
\reference{} Baldwin, J. A., et al. 1996, \apj, 468, L115
\reference{} Bautista, M. A., \& Pradhan, A. K. 1996, \aaps, 115, 551
\reference{} Bautista, M. A., \& Pradhan, A. K. 1997, \apj, in press
\reference{} Bautista, M. A., Pradhan, A. K., \& Osterbrock, D. E. 1994, \apj, 
432, 135
\reference{} Becklin, E. E., \& Beckwith, S., Gatley, I., Matthews, K., 
Neugebauer, G., Sarazin, C., \& Werner, M. W. 1976, \apj, 207, 770
\reference{} Black, J. H, \& van Dishoeck, E. F. 1987, \apj, 412, 322
\reference{} Bohlin, R. C., Savage, B. D., \& Drake, J. F. 1978, \apj, 224, 132
\reference{} Burton, M. G., Hollenbach, D. J., \& Tielens, A. G. G. M. 1990,
\apj, 365, 620
\reference{} DePoy, D. L, \& Pogge R. W. 1994, \apj, 433, 725
\reference{} DePoy, D. L, \& Shields, J. C. 1994, \apj, 422, 187
\reference{} Draine, B. T., \& Bertoldi, F. 1996, \apj, 468, 269 (DB96)
\reference{} Engelbracht, C. W., Rieke, M. J, Rieke, G. H., Kelly, D. M., 
\& Achtermann, J. M. 1997, \apj, in press
\reference{} Engelbracht, C. W., Rieke, M. J, \& Rieke, G. H. 1998, submitted
\reference{} Goldschmidt, O., \& Sternberg, A. 1995, \apj, 439, 256
\reference{} Hayashi, M., Hasegawa, T., Gatley, I., Garden, R., \& Kaifu, N. 
1985, \mnras, 215, 31
\reference{} Hogerheijde, M. R., Jansen, D. J., \& van Dishoeck, E. F. 1995,
\aap, 294, 792
\reference{} Hora, J. L., \& Latter, W. B. 1996, \apj, 461, 288
\reference{} Jansen, D. J., Spaans, M., Hogerheijde, M. R., \& van Dishoeck, 
E. F. 1995, \aap, 303, 541
\reference{} Johansson, S. 1978, Physica Scripta, 18, 217
\reference{} Keenan, F. P., Berrington, K. A., Burke, P. G., Zeippen, C. J.,
Le Dourneuf, M., \& Clegg, R. E. S. 1992, \apj, 384, 385
\reference{} Kulesa, C. Engelbracht, C. W., Ge, J., \& Rieke, G. H. 1998,
in prep.
\reference{} Livingston, W., \& Wallace, L. 1991, An atlas of the solar spectrum
in the infrared from 1850 to 9000 cm$^{-1}$, National Solar Observatory 
Technical Report 91-001
\reference{} Lowe, R. P., Moorehead, J. M., \& Wehlau, W. H. 1979, \apj,
228, 191
\reference{} Luhman, K. L., \& Rieke, G. H. 1996, \apj, 461, 298
\reference{} Luhman, M. L., Jaffe, D. T., Keller, L. D., \& Pak, S. 1994, \apj,
436, 185
\reference{} Luhman, M. L., Jaffe, D. T., Sternberg, A., Herrmann, F., \&
Poglitsch, A. 1997, \apj, 482, 298
\reference{} Lumsden, S. L., \& Puxley, P. J. 1996, \mnras, 281, 493
\reference{} Maiolino, R., Rieke, G. H., \& Rieke, M. J. 1996, \aj, 111, 537
\reference{} Marconi, A., Testi, L., Natta, A., \& Walmsley, C. M. 1997, \aap, 
in press
\reference{} Martin, P. G., Rubin, R. H., Ferland, G. J., Dufour, R. J., O'Dell,
C. R., Baldwin, J. A., Hester, J. J., \& Walter, D. K. 1996, \baas, 28, 106.01
\reference{} Martin, W. C. 1973, J. Phys. Chem. Ref. Data, 2, 257
\reference{} McMullen, J. P., Mundy, L. G., \& Blake, G. A. 1993, \apj, 405, 599
\reference{} O'Dell, C. R. 1994, \aaps, 216, 267
\reference{} Omodaka, T., Hayashi, M., Hasegawa, T., \& Hayashi, S. S. 1994, 
\apj, 430, 256
\reference{} Outred, M. 1978, J. Phys. Chem. Ref. Data, 7, 1
\reference{} Parmar, P. S., Lacy, J. H., \& Achtermann, J. M. 1991, \apj, 372,
L25
\reference{} Pogge, R. W., Owen, J. M., \& Atwood, B. 1992, \apj, 399, 147
\reference{} Rieke, G. H., \& Lebofsky, M. J. 1985, \apj, 288, 618
\reference{} Robbins, R. R. 1968, \apj, 151, 511
\reference{} Schmid-Burgk, J., G\"{u}sten, R., Mauersberger, R., Schulz, A., \& 
Wilson, T. L. 1990, \apj, 362, L25
\reference{} Sellgren, K., Tokunaga, A. T., \& Nakada, Y. 1990, \apj, 349, 120
\reference{} Smits, D. P. 1996, \mnras, 278, 683
\reference{} Storey, P. J., \& Hummer, D. G. 1995, \mnras, 272, 41
\reference{} Sternberg, A., \& Dalgarno, A. 1989, \apj, 338, 197
\reference{} Sternberg, A., \& Dalgarno, A. 1995, \apjs, 99, 565
\reference{} Sugar, J., \& Corliss, C. 1985, J. Phys. Chem. Ref. Data, 14, 
Suppl. 2
\reference{} Tauber, J. A., Tielens, A. G. G. M., Meixner, M., \& Goldsmith,
P. F. 1994, \apj, 422, 136
\reference{} Tielens, A. G. G. M., \& Hollenbach, D. 1985, \apj, 291, 722
\reference{} Tielens, A. G. G. M., Meixner, M. M., van der Werf, P. P., Bregman,
J., Tauber, J. A., Stutzki, J., \& Rank, D. 1993, Science, 262, 86
\reference{} Usuda, T., Sugai, H., Kawabata, H., Inoue, M. Y., Kataza, H., 
\& Tanaka, M. 1996, \apj, 464, 818
\reference{} van der Werf, P. P., Stutzki, J., Sternberg, A., \& Krabbe A.
1996, \aap, 313, 633
\reference{} Wilson, T. L., \& Mauersberger, R. 1991, \aap, 244, L33
\reference{} Williams, D., Thompson, C. L., Rieke, G. H, \& Montgomery, E. F.
1993, ProcSPIE, 1308, 482
\reference{} Wyrowski, F., Schilke, P., Hofner, P., \& Walmsley, C. M. 1997, 
\apj, 487, L171
\reference{} Ziurys, L. M., Martin, R. N., Pauls, T. A., \& Wilson, T. L. 1981,
\aap, 104, 288
\end{references}
\end{document}